\documentclass[12pt]{article}
\usepackage{amssymb}

\usepackage{bbm}
\usepackage{epsfig}
\usepackage{array}
\usepackage{float}
\usepackage{dsfont}
\usepackage{amstext}
\usepackage{psfrag}
\usepackage{a4}
\usepackage{a4wide}
\usepackage{setspace}

\begin{document}

\begin{flushright}
BA-09-02\\
%{\tt hep-ph/0708xxx}\\
%Aug, 2007
\end{flushright}
\vspace*{1.0cm}

\begin{center}
\baselineskip 20pt {\Large\bf  Hybrid Inflation Revisited in Light of WMAP5} \vspace{1cm}

{\large Mansoor Ur Rehman, Qaisar Shafi, Joshua R. Wickman} \vspace{.5cm}

{\baselineskip 20pt \it 
Bartol Research Institute, Department of Physics and Astronomy, \\
University of Delaware, Newark, Delaware 19716, USA \\

 }
\vspace{.5cm}

%\vspace{1.5cm} {\bf Abstract}
\end{center}

\begin{abstract}

We study the effects of including one-loop radiative corrections in a non-super-symmetric hybrid inflationary model. These corrections can arise from Yukawa couplings between the inflaton and right-handed neutrinos, and induce a maximum in the potential which admits hilltop-type solutions in addition to the standard hybrid solutions. We obtain a red-tilted spectral index $n_s$, consistent with WMAP5 data, for sub-Planckian values of the field. This is in contrast to the tree level hybrid analysis, in which a red-tilted spectrum is achieved only for trans-Planckian values of the field. Successful reheating is obtained at the end of the inflationary phase via conversion of the inflaton and waterfall fields into right-handed neutrinos, whose subsequent decay can explain the observed baryon asymmetry via leptogenesis.
\end{abstract}

It has been shown in the recent WMAP five year analysis (WMAP5) that tree level chaotic inflation driven by a quartic potential is excluded at the 99\% confidence level \cite{Komatsu:2008hk}. The tensor to scalar ratio $r$ in such a model turns out to be excessively large. However, this conclusion can be significantly altered provided one takes into account radiative corrections, especially those generated by any Yukawa couplings that may be present between the inflaton and fermion fields in the model. Such Yukawa couplings can be expected on general grounds, particularly in those inflation models containing right-handed neutrinos in which the observed baryon asymmetry is explained via type I leptogenesis \cite{Fukugita:1986hr}. It has been shown that radiatively improved quartic potential models with plausible values of the Yukawa couplings can yield values for $r$ and the scalar spectral index $n_s$ that lie well inside the 2-$\sigma$ bounds provided by WMAP5 \cite{NeferSenoguz:2008nn}.

In this paper, following Ref. \cite{NeferSenoguz:2008nn}, we carry out a similar analysis for a hybrid inflationary (HI) potential \cite{Linde:1991km}. In addition to providing a mechanism for generating the primordial baryon asymmetry, the extra couplings of the scalar fields in the model (especially the inflaton) to right-handed neutrinos also play an important role in how the scalars decay, and thus in the reheating of the universe. We will see that, as in the quartic case, the radiatively corrected model explores an expanded region of parameter space, and can result in better agreement with experimental data.

At its tree level, the HI potential can be written as \cite{Lazarides:2001zd}
\begin{equation}
V (\chi , \phi) =  \kappa^2 \left(M^2 - \frac{\chi^2}{4}\right)^2+\frac{m^2\phi^2}{2}+\frac{\lambda^2 \chi^2 \phi^2}{4},
\end{equation}
where $M$, $m$ are mass parameters and $\kappa$, $\lambda$ are dimensionless. The global minima of the potential lie at $(\langle \chi \rangle,\langle \phi \rangle)= (\pm 2M , 0) $. 
The effective mass squared of the field $\chi$ in the  $\chi = 0$ direction is $m^2_{\chi} = -\kappa^2 M^2 + \lambda^2\phi^2/2$. Thus, for $\phi > \phi_c = \frac{\sqrt{2}\kappa M}{\lambda}$ the only minimum of the potential $V (\chi , \phi)$ lies at $\chi = 0$. In this region the HI potential takes the form
\begin{equation}
V (\phi) =  V_0 + \frac{m^2\phi^2}{2} = V_0 \left[1+ \widetilde{\phi}^2 \right],
\label{trpot}
\end{equation}
where  $\widetilde{\phi} \equiv \frac{m\,\phi}{\sqrt{2\,V_0}}$, and $V_0 = \kappa^2 M^4$ is the constant vacuum energy term. The second term in Eq. (\ref{trpot}) provides a non-zero slope in the otherwise flat potential, and the system can inflate as it rolls down the $\chi = 0$ valley. Upon reaching $\phi=\phi_c$, the minimum in the $\chi$ direction becomes a maximum, and inflation ends abruptly as the system rapidly falls into the global minimum. This scenario is termed ``hybrid" because the vacuum energy density $V_0$ is provided by the waterfall field $\chi$, while $\phi$ is the slowly rolling inflaton field.

Before considering radiative corrections, it is appropriate to discuss the predictions of the tree level hybrid inflationary (TLHI) model in comparison with WMAP5. The slow-roll parameters for the TLHI potential are given as
\begin{eqnarray}
\epsilon &=& \frac{m_P^2}{2}\left( \frac{\partial_\phi V}{V}\right)^2 =  
\frac{\eta_0}{4}\left( \frac{\partial_{\widetilde{\phi}} V}{V}\right)^2 =
\frac{\eta_0 \,\widetilde{\phi}^2}{\left(\widetilde{\phi}^2 +1\right)^2}, \\
 \eta &=& m_P^2 \left( \frac{\partial_\phi^2 V}{V} \right) =
 \frac{\eta_0}{2} \left( \frac{\partial_{\widetilde{\phi}}^2 V}{V} \right)=\frac{\eta_0}{\widetilde{\phi}^2 +1},
\end{eqnarray}
where $\eta_0 = \eta (\widetilde{\phi} = 0)= \frac{m^2\, m_P^2}{V_0}$, and $m_P\simeq2.4\times10^{18}$ GeV is the reduced Planck mass. The number of e-foldings after the comoving scale $l_0=2\pi /k_0$ has crossed the horizon is
given by
\begin{eqnarray}\label{efold1}
N_0 &=& \frac{1}{m_P^2}\int_{\phi_c}^{\phi_0} \left(\frac{V}{\partial_{\phi} V} \right) d\phi
=\frac{2}{\eta_0}\int_{\widetilde{\phi}_c}^{ \widetilde{\phi}_0} \left(\frac{V}{\partial_{ \widetilde{\phi}} V} \right) d \widetilde{\phi} = \frac{1}{2\,\eta_0} \left( \widetilde{\phi}_0^2-\widetilde{\phi}_c^2 + \text{ln } \widetilde{\phi}_0^2 / \widetilde{\phi}_c^2\right), 
\end{eqnarray}
where $\phi_0$ is the value of the field when the scale corresponding to $k_0$ exits the horizon, and $\phi_c = \sqrt{\frac{2\,\kappa}{\lambda^2}}\,V_0^{1/4}$ (or $\widetilde{\phi}_c = \sqrt{\frac{\eta_0}{2}} \left( \phi_c / m_P \right) $), with $\kappa \sim \lambda \sim 10^{-3}$, is the value of the field at the end of inflation. In Eq. (\ref{efold1}), we may eliminate $\eta_0$ in favor of $V_0$ using the curvature perturbation constraint
\begin{eqnarray}
\Delta_\mathcal{R} = \frac{1}{2\sqrt{3}\,\pi \,m_{P}^3}\frac{V^{3/2}}{\vert\partial_{\phi} V\vert}\vert_{\phi = \phi_0}
&=& \frac{1}{\sqrt{6\,\eta_0}\,\pi \,m_{P}^2}\frac{V^{3/2}}{\vert\partial_{\widetilde{\phi}} V\vert}\vert_{\widetilde{\phi} = \widetilde{\phi}_0} \label{cpc} \\
\Rightarrow \; \eta_0 &=& \frac{\left(1+ \widetilde{\phi}_0^2 \right)^3}{ 24\,\pi^2 \,\Delta_\mathcal{R}^2 \, \widetilde{\phi}_0^2  } \left( \frac{V_0}{m_P^4} \right) .
\end{eqnarray}
In our calculations, we will use the value $\Delta_\mathcal{R}(k_0)=4.91\times10^{-5}$ obtained by the recent WMAP5 analysis for $k_0=0.002$ Mpc$^{-1}$ \cite{Komatsu:2008hk}. To leading order, the spectral index $n_{s}$ and the tensor-to-scalar ratio $r$ are given by
\begin{eqnarray}
n_{s} &\simeq & 1 - 6 \epsilon + 2 \eta = 1 - 4\,\eta_0\, \frac{\left( \widetilde{\phi}_0^2 -1/2 \right)}{\left(\widetilde{\phi}_0^2+1\right)^2 },\label{ns}\\
r  &\simeq& 16 \epsilon = \frac{16\, \eta_0 \,\widetilde{\phi}_0^2}{\left(\widetilde{\phi}_0^2 +1\right)^2} = 4 (1-n_s)\frac{\widetilde{\phi}_0^2}{\widetilde{\phi}_0^2 -1/2}.
\end{eqnarray}
\begin{figure}[t] 
\begin{center}
\includegraphics[angle=0, width=12cm]{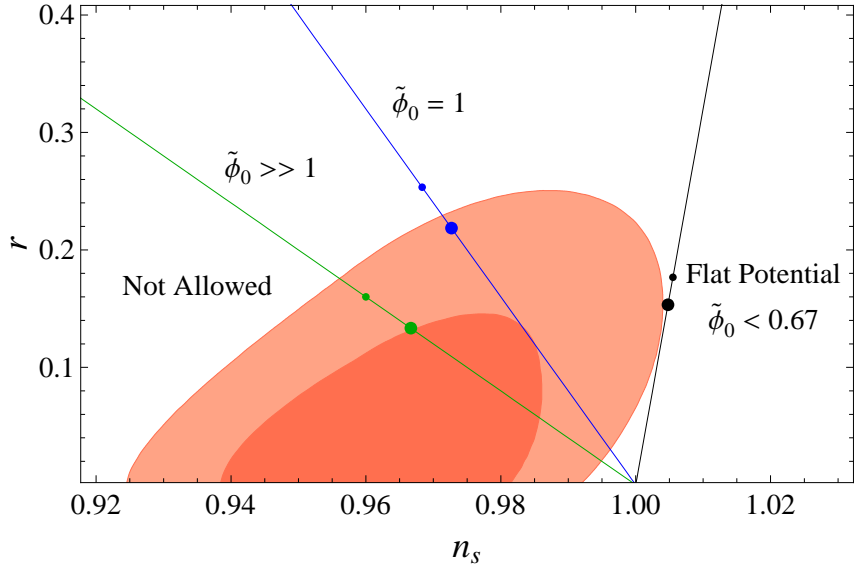} 
\caption{$r$ vs $n_s$ for the tree level hybrid inflationary potential, shown together with the WMAP5+BAO+SN contours (68\% and 95\% confidence levels) \cite{Komatsu:2008hk}. Here the small (large) dots correspond to $N_0 = 50$ ($N_0 = 60$).} \label{rvsns1}
\end{center}
\end{figure}
Using the above equations, we show in Fig. \ref{rvsns1} the predictions of the TLHI model along with the WMAP5 1-$\sigma$ and 2-$\sigma$ bounds, and specify the range in which $50<N_0<60$. The flat potential regime ($\widetilde{\phi_0}\ll 1$) lies outside the 2-$\sigma$ bound, corresponding to a blue spectral index. A red-tilted spectral index is obtained for $\widetilde{\phi}_0 > 1/\sqrt{2}\simeq 0.71$, and for even larger values of $\widetilde{\phi}_0$ this result reduces to the quadratic potential prediction $r \simeq 4 (1-n_s)$. It is interesting to study the value of the field $\phi$ compared to the reduced Planck mass $m_P$ (i.e., $\phi_P \equiv \frac{\phi_0}{m_P}$) for which the sprectral index is red-tilted. With $\widetilde{\phi_0} > 1/\sqrt{2}$ and the slow-roll approximation  $(\eta,\epsilon) \ll 1 $ (which implies $\eta_0\ll 1$), the blue spectrum can only be avoided for trans-Planckian values of the field $\phi_P = \sqrt{\frac{2}{\eta_0}}\, \widetilde{\phi_0} > \sqrt{\frac{1}{\eta_0}} \gg 1$. For a more exact treatment obtaining a red spectrum in the TLHI model, see the recent paper in Ref. \cite{Clesse:2008pf}.  With these tree level results in mind, we would now like to examine the situation after including  radiative corrections in the HI potential. 

Consider the following Lagrangian density
\begin{eqnarray}
{\cal L} &=&  \frac12\partial^{\mu}\phi_B\partial_{\mu}\phi_B + \frac12\partial^{\mu}\chi_B\partial_{\mu}\chi_B + \frac{i}{2}\bar{N}\gamma^{\mu}\partial_{\mu}N -\kappa^2 \left(M^2 - \frac{\chi_B^2}{4}\right)^2  \nonumber \\
&& -\frac{m_B^2\phi_B^2}{2}-\frac{\lambda_B^2 \chi_B^2 \phi_B^2}{4}
-\frac{1}{2}y_B \phi_B \bar{N}N-\frac{1}{2}Y_B \chi_B \bar{N}N-\frac{1}{2}m_N \bar{N}N,\label{lagrangian}
\end{eqnarray}
where the subscript `B' denotes bare quantities. To keep the discussion as simple as possible, we have introduced a single Yukawa coupling involving $N$ and each of $\phi$ and $\chi$. In a more realistic scenario, successful leptogenesis requires at least two right-handed neutrinos. Since $\chi=0$ during inflation, the interaction between $N$ and $\chi$ has no effect in this regime. After inflation, however, $N$ acquires a contribution $M_N\simeq Y \langle\chi\rangle$ to its mass in addition to the bare mass $m_N$. Also, oscillations of $\chi$ affect the way in which the system reheats after inflation, as we will see later when we discuss the reheating phase.

The inflationary potential including one-loop corrections, in terms of renormalized quantities, is given by 
\begin{equation}
V = \kappa^2 \left(M^2 - \frac{\chi^2}{4}\right)^2+\frac{m^2\phi^2}{2}+\frac{\lambda^2 \chi^2 \phi^2}{4}+V_{\rm loop},
\end{equation}
where $V_{\rm loop}$ is the one-loop correction to the tree level potential. In the $\chi = 0$ direction, $V_{\rm loop}$ can be written as \cite{Coleman:1973jx}
\begin{eqnarray}
V_{\text{loop}} &=& \frac{1}{64\pi^2}\left[m^4\text{ ln} \left( \frac{m^2}{\mu^2} \right) + \frac{\lambda^4}{4}\left( \phi^2 - \phi_c^2 \right)^2 \text{ ln} \left( \frac{ \frac{\lambda^2}{2} \left(\phi^2 - \phi_c^2 \right) }{\mu^2} \right)  \right. \nonumber \\
&&  \left. - 2 \left(m_N + y \phi \right)^4 \text{ ln} \left( \frac{m_N + y \phi}{\mu} \right)^2 \right] .
\end{eqnarray}
During inflation, $\phi$ is always larger than $\phi_c$, therefore 
for $y \phi_c \gg (m_N,m)$ and $y \gtrsim \frac{\lambda}{\sqrt{2}}$, the above potential reduces to 
\begin{equation}
V_{\text{loop}} = -A \phi ^4 \text{ ln} \left( \frac{y \phi}{\mu} \right) , \text{ with } A = \frac{y^4}{16\pi^2}.
\end{equation}
In order to ensure that the log factor is always positive during inflation, it is convenient to set the renormalization scale at $\mu=y\,\phi_c$. Then, the complete radiatively-corrected hybrid inflationary (RCHI) potential in the $\chi = 0$ direction reduces to the form
\begin{equation}\label{V2}
V =  V_0 +\frac{m^2\phi^2}{2}-A_{\phi} \, \phi^4 = V_0 \left[1+ \widetilde{\phi}^2 - \widetilde{A}_{\phi}\, \widetilde{\phi}^4 \right] ,
\end{equation}
where $\widetilde{A}_{\phi} = \frac{4\,A_{\phi}}{\eta_0^2\,(V_0/m_P^4)}$ and $A_{\phi} = A\text{ ln}\left( \frac{\phi}{\phi_c}\right) $. In the following calculations, we will approximate $\widetilde{A}_{\phi}$ to be independent of $\phi$. The slow-roll parameters in this case are given by
\begin{eqnarray}
\epsilon = \frac{\eta_0 \left( \widetilde{\phi} - 2\,\widetilde{A}_{\phi}\, \widetilde{\phi}^3 \right)^2 }{\left(1+ \widetilde{\phi}^2 - \widetilde{A}_{\phi}\, \widetilde{\phi}^4 \right)^2}, \;\;\;\;
 \eta =\frac{\eta_0\left( 1 - 6\,\widetilde{A}_{\phi}\, \widetilde{\phi}^2 \right) }{1+ \widetilde{\phi}^2 - \widetilde{A}_{\phi}\, \widetilde{\phi}^4 }.
\end{eqnarray}
Using Eq. (\ref{cpc}) and solving for $\eta_0$ as before, we find
\begin{equation}
\eta_0  = \frac{ \left(1+ \widetilde{\phi}_0^2 - \widetilde{A}_{\phi}\, \widetilde{\phi}_0^4 \right)^3 }{24\,\pi^2 \, \Delta_\mathcal{R}^2  \, \widetilde{\phi}_0^2 \left(1 - 2\widetilde{A}_{\phi}\, \widetilde{\phi}_0^2 \right)^2}\left( \frac{V_0}{m_P^4} \right).
\end{equation}
The number of e-foldings in the RCHI model can be calculated as
\begin{eqnarray}
N_0 = \frac{1}{2\,\eta_0} \left(\frac{\widetilde{\phi}_0^2-\widetilde{\phi}_c^2}{2}  + \text{ln } \widetilde{\phi}_0^2 / \widetilde{\phi}_c^2-\frac{\left( 1+4\,\widetilde{A}_{\phi} \right) \text{ln }\left[ \frac{1-2\,\widetilde{A}_{\phi}\widetilde{\phi}_0^2}{1-2\,\widetilde{A}_{\phi}\widetilde{\phi}_c^2}\right] }{4\,\widetilde{A}_{\phi}} \right).
\end{eqnarray}
In order to ensure that the potential remains bounded during inflation, we take $\widetilde{\phi}_c < \widetilde{\phi}_0 < \widetilde{\phi}_M = \frac{1}{\sqrt{2\,\widetilde{A}_{\phi}}}$, where $\widetilde{\phi}_M$ is field value at the maximum of the potential. This maximum introduces hilltop-type solutions \cite{Boubekeur:2005zm}, to which we now turn our discussion.

In its approximate form, the potential in Eq. (\ref{V2}) has previously been analyzed only for sub-Planckian hilltop-type solutions \cite{Kohri:2007gq}, for which inflation begins in a region where the potential is concave downward. In general, the RCHI model can lead to both hilltop and non-hilltop solutions (see Ref. \cite{NeferSenoguz:2008nn} for a similar analysis in the case of radiatively-corrected quadratic and quartic potentials). In order to study both types of solutions, it will be convenient to define the quantities $f\equiv\widetilde{\phi}_0 /\widetilde{\phi}_M$, $f_c\equiv\widetilde{\phi}_c /\widetilde{\phi}_M$  and $f_1\equiv\sqrt{2\,\widetilde{A}_{\phi}}$. As we will see, for the same values of $\widetilde{A}_{\phi}$ and $V_0$, we can always distinguish two separate branches of solutions, one with large $f$ values and another with small $f$ values. To facilitate our discussion, we can rewrite the number of e-foldings in the form
\begin{eqnarray}
N_0 &=& \frac{1 }{2 \eta_0} \times \left( \frac{f^2-f^2_c}{2\,f_1^2}+\text{ln}\frac{f^2}{f_c^2} + \frac{\left( f_1^2 +1/2 \right)}{f_1^2} \text{ln} \frac{1-f_c^2}{1-f^2}\right), \\
\text{ where }  \nonumber \\ 
f_c &=& f_1 \, \left( \frac{V_0^{1/4}}{m_P} \right)  \, \sqrt{\frac{2\, \eta_0}{y}}, \;\;\;\; \eta_0 = \frac{V_0/m_P^4}{24\, \pi^2 \, \Delta_\mathcal{R}^2} \left( \frac{\left(1+\left(f/f_1 \right)^2 \left( 1-f^2/2 \right) \right)^3}{\left(f/f_1 \right)^2\left( 1-f^2 \right)^2 } \right).
\end{eqnarray}

\begin{figure}[t] 
\begin{center}
\includegraphics[angle=0, width=17cm]{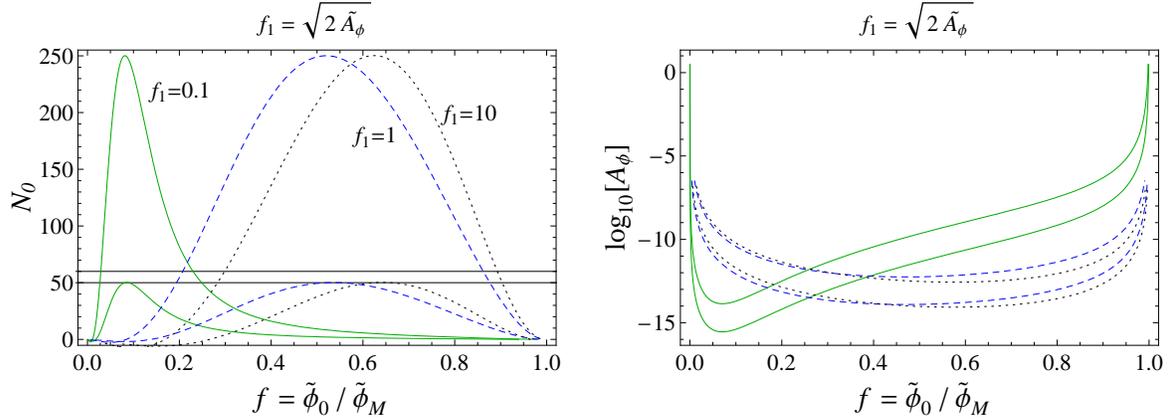} 
\caption{$N_0$ and log$_{10}\left[ A_{\phi} \right]$  vs $f$ for $f_1 = 0.1, 1$ and $10$ with $V_0^{1/4} = (10^{16.15}-10^{16.29})$ GeV, $(10^{16.08}-10^{16.22})$ GeV and $(10^{15.61}-10^{15.74})$ GeV respectively. We obtain two solutions $(n_s,r)$ for each value of $A_{\phi}$ and $V_0$.} \label{n0vsf}
\end{center}
\end{figure}

 In order to discuss the two branches qualitatively, we note that the factor inside parentheses in $N_0$ is slowly varying over the entire range of $f$. Therefore, it is sufficient to consider only the contribution of the $\eta_0^{-1}$ factor in $N_0$. For large values of $f_1$, $\eta_0^{-1}$ reduces to the form $f^2(1-f^2)^2$, with a maximum at $f=1/\sqrt{3}\sim 0.58$. As we move away from this maximum, we obtain the same value of $N_0$ for each of two values of $f$, one smaller and one larger than $1/\sqrt{3}$. These large-$f$ and small-$f$ branches can be seen in Fig. \ref{n0vsf}. As $f_1$ is lowered, the maximum of $N_0$ moves toward smaller values of $f$, asymptotically tending toward $f=f_1/\sqrt{2}$. 

Let us discuss in greater detail the large $f_1$ limit, which corresponds to the ``flat potential regime" (i.e. $\widetilde{\phi_0}=f/f_1\ll 1$). In this limit, the number of e-foldings $N_0$ reduces to
\begin{eqnarray}
N_0 &\simeq& \frac{12\, \pi^2 \, \Delta_\mathcal{R}^2}{V_0/m_P^4} \left(  \frac{f^2\left( 1-f^2 \right)^2}{f_1^2} \right)  \text{ln}\frac{f_c^2(1-f_c^2)}{f^2(1-f^2)},
\end{eqnarray}
with 
\begin{eqnarray}
f_c \simeq  \sqrt{\frac{1}{12\,y\, \pi^2 \, \Delta_\mathcal{R}^2}} \left( \frac{V_0^{1/4}}{m_P} \right)^3  \left( \frac{f_1^2}{f^2\left( 1-f^2 \right) } \right)\,f.
\end{eqnarray}
An upper bound on $V_0$ can be found by considering the point at which the two branches meet for a given value of $f_1$. Owing to the inverse dependence of $N_0$ on $V_0$, the number of e-foldings at the maximum is shifted upward as $V_0$ is decreased away from its limiting value, and the two branches diverge from one another in the range corresponding to realistic values of $N_0$. The basic condition $f>f_c$ (or $N_0>0$) leads to the bound $V_0^{1/4} \lesssim \frac{3 \times 10^{16}}{f_1^{2/3}}$ GeV. In order to obtain a realistic number of e-foldings, $V_0$ must obey a somewhat more stringent bound.

As mentioned earlier, hilltop solutions are defined as having a concave downward curvature of the potential when inflation begins. Therefore, only those solutions which satisfy $f\gtrsim f_i= 1/\sqrt{3} \simeq 0.58$ will be called hilltop solutions, where $f_i$ is the value of $f$ at the point of inflection.

\begin{figure}[t] 
\begin{center}
\includegraphics[angle=0, width=17cm]{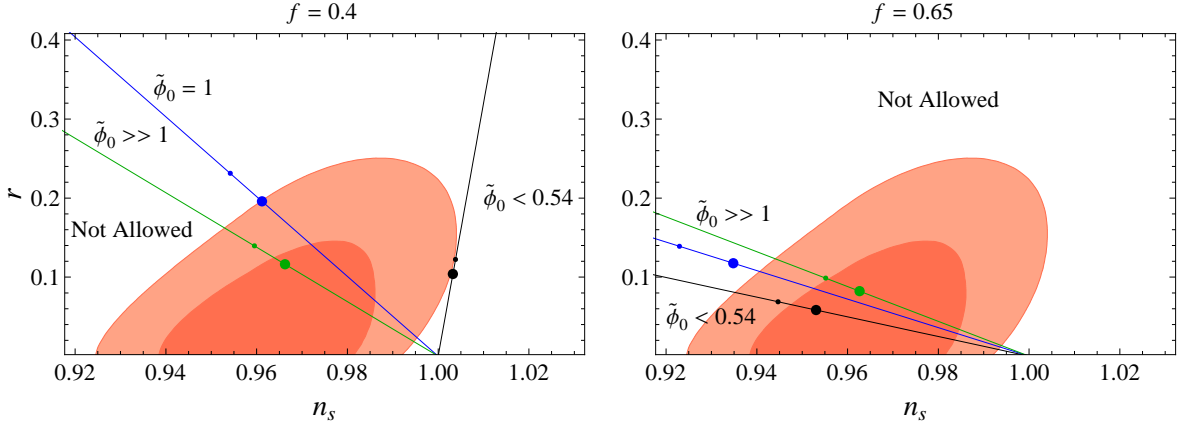} 
\caption{$r$ vs $n_s$ for radiatively-corrected hybrid inflation, with $f =  0.4$ and $f =  0.65$. Here, small (large) dots correspond to $N_0 = 50$ ($N_0 = 60$). Compared to Fig. \ref{rvsns1}, smaller values of $\widetilde{\phi_0}$ fall within the WMAP5 bounds. In addition, the allowed and disallowed regions are exchanged in going from a non-hilltop ($f<f_i$) to a hilltop ($f>f_i$) solution.} \label{rvsns2}
\end{center}
\end{figure}
 
The spectral index $n_s$  and the tensor-to-scalar ratio $r$ for the RCHI model are given as
\begin{eqnarray}
n_s &=& 1 + 2\,\eta_0\, \frac{\left( 1 -2(1+3\widetilde{A}_{\phi})\widetilde{\phi}_0^2 +5\widetilde{A}_{\phi}\widetilde{\phi}_0^4 - 6\widetilde{A}_{\phi}^2 \widetilde{\phi}_0^6 \right)}{\left(1+ \widetilde{\phi}_0^2 - \widetilde{A}_{\phi}\, \widetilde{\phi}_0^4 \right)^2 },\\
r &=& \frac{16\,\eta_0 \left( \widetilde{\phi}_0 - 2\,\widetilde{A}_{\phi}\, \widetilde{\phi}_0^3 \right)^2 }{\left(1+ \widetilde{\phi}_0^2 - \widetilde{A}_{\phi}\, \widetilde{\phi}_0^4 \right)^2} = 4 (1-n_s)\frac{\widetilde{\phi}_0^2 \, \left( 1-f^2\right)^2 }{\widetilde{\phi}_0^2\left(1-\frac{5}{4}f^2+\frac{3}{4}f^4 \right)-\frac{1}{2}\left( 1-\frac{3}{2}f^2\right) }\label{r1loop}.
\end{eqnarray}
The spectral index becomes unity when $\widetilde{\phi}_0$ acquires the value
\begin{equation}\label{phi0ns1}
\widetilde{\phi}_0 = \frac{1}{\sqrt{2}}\left( \frac{\sqrt{1-3f^2}}{\sqrt{1-\frac{5}{4}f^2+\frac{3}{4}f^4}}\right) ,
\end{equation}
which reduces to the tree level result $\widetilde{\phi_0}=\frac{1}{\sqrt{2}}$ in the $ f \rightarrow 0$ limit. For $\widetilde{\phi_0}$ values larger (smaller) than in Eq. (\ref{phi0ns1}), we obtain a red-tilted (blue-tilted) spectral index if $f < f_i$. For $f > f_i$, $\eta$ becomes negative and we always have a red-tilted spectral index.

\begin{figure}[t] 
\begin{center}
\includegraphics[angle=0, width=17cm]{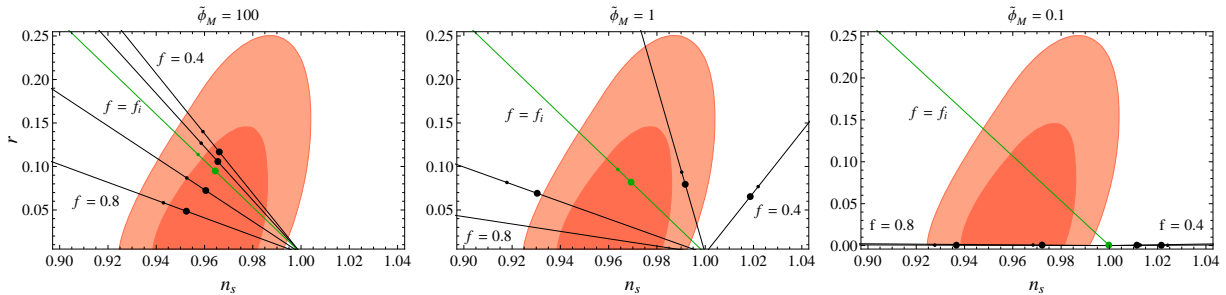} 
\caption{$r$ vs $n_s$ for radiatively-corrected hybrid inflation, with  $\widetilde{\phi}_M =  100, 1 $ and $0.1$ for $f = 0.4,0.5,f_i,0.7$ and $0.8$. Large values of $\widetilde{\phi}_M$ result in appreciable values of $r$ for a realistic number of e-foldings.} \label{rvsns3}
\end{center}
\end{figure}

For small values of $f$, Eq. (\ref{r1loop}) reduces to the tree level result $r \simeq 4 (1-n_s)\frac{\widetilde{\phi}_0^2}{\widetilde{\phi}_0^2 -1/2}$.  Similarly, for large values of $\widetilde{\phi}_0$ we find $r \simeq 4 (1-n_s)\frac{\left( 1-f^2\right)^2 }{\left(1-\frac{5}{4}f^2+\frac{3}{4}f^4 \right) }$, which reduces to the quadratic prediction $r \simeq 4 (1-n_s)$ in the small $f$ limit (see Figs. \ref{rvsns2} and \ref{rvsns3}). Thus we see that in order for the radiative corrections to produce a reasonably large contribution, $\widetilde{\phi}_0$ should be chosen close to $\widetilde{\phi}_M$. This choice in turn shifts the ``flat potential regime" toward the inside of the 2-$\sigma$ bound of the WMAP+BAO+SN data, as can be seen by comparing Figs. \ref{rvsns1} and \ref{rvsns2}. A larger choice of $f$ leads to a greater portion of the ``flat" region inside the bounds. Additionally, Fig. \ref{rvsns2} shows that for $f < f_i$ the allowed regime lies above the $\widetilde{\phi_0}=100$ line, whereas for $f > f_i$ the allowed regime lies below this line. As noted earlier, for $f > f_i$, $\eta$ becomes negative and both $\epsilon$ and $\eta$ tend to drive $n_s$ ($\simeq1-6\,\epsilon +2\,\eta$) below 1 as $\eta_0$ increases. In contrast, for $f < f_i$, $\eta$ is always positive and competes with $\epsilon$, allowing $n_s$ values above or somewhat below 1.

In our discussion so far, we have suppressed the $\phi$ dependence of $A_{\phi}$ in order to obtain analytically tractable expressions.  Next, we employ numerical calculations to examine the RCHI model in such a way that this dependence can be taken into account.  In these calculations, we use the next-to-leading order expressions for $n_s$, $r$, and $\Delta_\mathcal{R}$ for added precision \cite{Stewart:1993bc}.  Having already explored the relative behavior of the quadratic and vacuum terms, we will use the form of the potential Eq. (\ref{V2}) written in terms of $\phi$ rather than $\widetilde{\phi}$ in order to more directly probe the parameters $m^2$ and $V_0$.

As discussed earlier, inflation ends via a waterfall effect at the critical value $\phi_c$.  It is worth noting that inflation can end due to the breakdown of the slow-roll approximation, before $\phi$ reaches this value; however, in our calculations below, $|\eta|\lesssim 10^{-2}$ and $\epsilon\lesssim 10^{-5}$ when evaluated at $\phi_c$, and so the slow-roll relations are still valid when the waterfall is induced.

\begin{table}[!b]
\resizebox{16.0 cm}{!}{
%\begin{displaymath}
$\begin{array}{||c|c|c|c|c|c|c|c|c|c|c||}
\hline
 f & V_0^{1/4}\, (\rm{GeV}) & A\, (10^{-13}) & M\, (\rm{GeV}) & m\, (\rm{GeV}) & f_c & n_s & r\, (10^{-3}) & m_{\phi}\, (10^{14}\, \rm{GeV}) & T_r\, (10^{12}\, \rm{GeV}) & N_0 \\
\hline\hline
\multicolumn{11}{||c||}{\phi_P = 0.25 }\\
\hline
 0.65 & 3.696\times 10^{15} & 1.164 & 5.744\times 10^{16} & 9.04\times 10^{11} & 0.1735 & 0.9864 & 0.149 & 1.68 & 1.45 & 56.30
   \\
 0.70 & 3.584\times 10^{15} & 1.282 & 5.503\times 10^{16} & 8.74\times 10^{11} & 0.1789 & 0.9717 & 0.132 & 1.65 & 1.47 & 56.29
   \\
 0.75 & 3.434\times 10^{15} & 1.336 & 5.245\times 10^{16} & 8.29\times 10^{11} & 0.1828 & 0.9536 & 0.111 & 1.59 & 1.46 & 56.25
   \\
 0.80 & 3.235\times 10^{15} & 1.296 & 4.960\times 10^{16} & 7.63\times 10^{11} & 0.1843 & 0.9312 & 0.0870 & 1.49 & 1.41 & 56.19
   \\
\hline\hline
\multicolumn{11}{||c||}{\phi_P = 1}\\
\hline
 0.60 & 8.873\times 10^{15} & 2.134 & 1.278\times 10^{17} & 6.13\times 10^{12} & 0.08907 & 0.9930 & 5.02 & 4.36 & 2.72 & 57.10
   \\
 0.65 & 8.603\times 10^{15} & 2.333 & 1.225\times 10^{17} & 5.87\times 10^{12} & 0.09252 & 0.9768 & 4.43 & 4.27 & 2.75 & 57.08
   \\
 0.70 & 8.262\times 10^{15} & 2.430 & 1.171\times 10^{17} & 5.53\times 10^{12} & 0.09521 & 0.9575 & 3.77 & 4.12 & 2.73 & 57.04
   \\
 0.75 & 7.836\times 10^{15} & 2.387 & 1.113\times 10^{17} & 5.10\times 10^{12} & 0.09696 & 0.9346 & 3.04 & 3.90 & 2.65 & 56.99
   \\
\hline\hline
\multicolumn{11}{||c||}{\phi_P = 2.5}\\
\hline
 0.55 & 1.482\times 10^{16} & 2.280 & 2.116\times 10^{17} & 1.88\times 10^{13} & 0.05409 & 0.9921 & 43.1 & 7.34 & 3.59 & 57.55
   \\
 0.60 & 1.443\times 10^{16} & 2.549 & 2.032\times 10^{17} & 1.81\times 10^{13} & 0.05665 & 0.9783 & 38.7 & 7.24 & 3.67 & 57.53
   \\
 0.65 & 1.394\times 10^{16} & 2.728 & 1.947\times 10^{17} & 1.72\times 10^{13} & 0.05880 & 0.9619 & 33.7 & 7.06 & 3.68 & 57.50
   \\
 0.70 & 1.334\times 10^{16} & 2.779 & 1.859\times 10^{17} & 1.60\times 10^{13} & 0.06046 & 0.9427 & 28.2 & 6.77 & 3.62 & 57.48 \\
\hline
\end{array}$
%\end{displaymath}
\par}
\caption{Predicted values of various inflationary parameters in the radiatively-corrected hybrid model. Here we show only those values falling within the WMAP5+BAO+SN 2-$\sigma$ bounds (see Fig. \ref{nsr_num}) for each of the three cases $\phi_P \equiv \frac{\phi_0}{m_P}= $ 0.25, 1, and 2.5, which correspond primarily to hilltop solutions ($f\gtrsim 0.6$). These values correspond to the choice $\phi_c \simeq 2 V_0^{\frac{1}{4}}/\sqrt{y}$ in Eq.(\ref{phic}).}\label{table1}
\end{table}

In terms of the reheat temperature $T_r$, the number of observable e-foldings may be written as \cite{Liddle:2003as}
\begin{eqnarray}\label{efold2}
N_0&\simeq& 53+\frac{2}{3}\, \text{ln}\left[ \frac{V(\phi_0)^{1/4}}{10^{15} \text{ GeV}}\right] +\frac{1}{3}\, \text{ln}\left[ \frac{T_r}{10^{9} \text{ GeV}}\right], 
\end{eqnarray}
where $T_r \simeq \left[ \frac{30/\,g_*}{2\pi^3(1+w_{reh})(5-3w_{reh})}\right]^{1/4}\sqrt{\Gamma\,m_P}$ \cite{Martin:2006rs}, and $\Gamma$ is some appropriate decay width. Two sources of primordial $N$ production contribute to reheating, resulting from the oscillations of $\phi$ and $\chi$ about their respective minima. The decay rates of these two processes are given by
\begin{equation}\label{decayrates}
\Gamma_{\chi \rightarrow N\,N} = \frac{Y^2 \,m_{\chi}}{8\pi}, \, \, \, \, \Gamma_{\phi \rightarrow N\,N} = \frac{y^2 \,m_{\phi}}{8\pi},
\end{equation}
where $Y\simeq M_N /\langle\chi\rangle$, and the scalar field masses are given as
\begin{equation}\label{masses}
m_{\chi} = \sqrt{2}\,\kappa M, \, \, \, \, m_{\phi} = \sqrt{m^2+2\left( \lambda M\right)^2 }.
\end{equation}
To see how these two contributions compare, consider the ratio of their decay rates
\begin{equation}\label{decayratio}
\frac{\Gamma_{\phi \rightarrow N\,N}}{\Gamma_{\chi \rightarrow N\,N}} = \left( \frac{y}{Y}\right)^2 \frac{m_{\phi}}{m_{\chi}}.
\end{equation}
To simplify the analysis, we will take $\lambda\sim\kappa\sim y\sim Y$. Then, for $m\ll M$, we obtain $m_{\phi}\sim m_{\chi}$ and the ratio of the decay rates in Eq. (\ref{decayratio}) is of order unity. Indeed, under these assumptions, it turns out that all relevant decay widths are of the same order, and so we may approximate the total decay width appearing in $T_r$ as $\Gamma\sim\Gamma_{\phi\rightarrow N\,N}$. Then, using $w_{reh}=0$ for matter dominant reheating and taking $g_*\simeq 106$, the reheat temperature becomes $T_r \simeq 0.035\,y \sqrt{m_{\phi}\,m_P}$.

To proceed further, we may eliminate $m^2$ in favor of the field value at the local maximum $\phi_M$ induced by the radiative correction term.  Assuming this maximum is the only extremum other than the minimum at the origin, we can write
\begin{equation}
m^2 = A\phi_M^2 \left(1+4\ln\frac{\phi_M}{\phi_c} \right).
\end{equation}
Using the approximations above, we may simplify the expression for the critical value of the inflaton field:
\begin{equation}\label{phic}
\phi_c = \sqrt{\frac{2\,\kappa}{\lambda^2}}\,V_0^{1/4}\sim \frac{V_0^{\frac{1}{4}}}{\sqrt{y}}.
\end{equation}
Note that this expression depends only on $V_0$ and $A$ (via $y$).

We are interested in comparing sub-Planckian and trans-Planckian inflation.  We thus consider three values of the inflaton at the start of inflation: $\phi_P = 0.25 $, $\phi_P = 1$ and $\phi_P = 2.5 $.  In each of these cases, if the ratio $f$ is fixed, $\phi_M$ is known and the inflationary potential $V(\phi)$ is specified in terms of $V_0$ and $A$.

To perform our numerical calculations, we fix a value of $f$ and scan over values of $V_0$ until the number of e-foldings $N_0$ given by an integral similar to Eq. (\ref{efold1}) matches its value as given by thermal considerations, Eq. (\ref{efold2}).  For each $(\phi_0,f,V_0)$, the value of $A$ can then be calculated by setting the curvature perturbation equal to its WMAP5 value.  The results of these calculations are displayed in Table \ref{table1}.

\begin{figure}[t] 
\begin{center}
\includegraphics[angle=0, width=16cm]{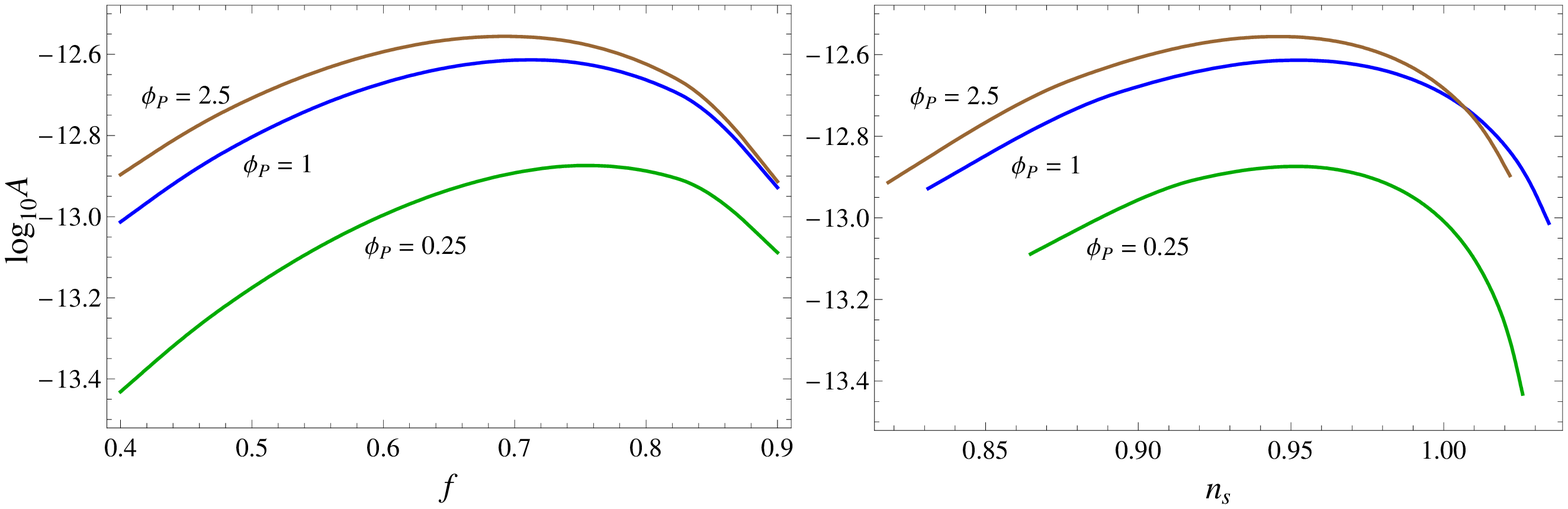} 
\caption{$\log_{10}A$ vs. $f$ and $n_s$ for the cases $\phi_P \equiv \frac{\phi_0}{m_P}= $ 0.25, 1, and 2.5. Two solutions of each of $f$, $n_s$ exist for a single value of $A$ (see Fig. \ref{n0vsf}). The region falling within the WMAP5 bounds mainly leads to two hilltop solutions.} \label{flogA}
\end{center}
\end{figure}

The RCHI model yields values of the reheat temperature on the order of $T_r\sim 10^{12}$ GeV, which is substantially larger than the range allowed by supersymmetric models of inflation, $T_r\sim 10^6$-$10^9$ GeV \cite{Senoguz:2004vu}.  This range is formulated based on the gravitino constraint, which is peculiar to susy models and does not pose a danger to the non-susy model that we currently consider.  Table \ref{table1} also shows that, in order to obtain a spectral index near the WMAP5 central value, the RCHI model predicts $V_0^{1/4}\sim 10^{15}$-$10^{16}$ GeV.  A similar range of preferred $V_0^{1/4}$ values has recently been realized in other models such as susy hybrid inflation \cite{Dvali:1994ms} and non-susy Coleman-Weinberg and Higgs inflation \cite{Rehman:2008qs}.  In the non-susy models, this range was also seen to be associated with proton decay, predicting a lifetime of $10^{34}$-$10^{38}$ years. In addition, such models predict a reheat temperature 4-6 orders of magnitude lower than the RCHI model, due in part to a Yukawa coupling on the order of $10^{-6}$. In contrast, the Yukawa coupling involving $\phi$ in the RCHI model is linked to the coefficient $A\sim 10^{-13}$, implying $y\sim 10^{-3}$.

\begin{figure}[t] 
\begin{center}
\includegraphics[angle=0, width=13cm]{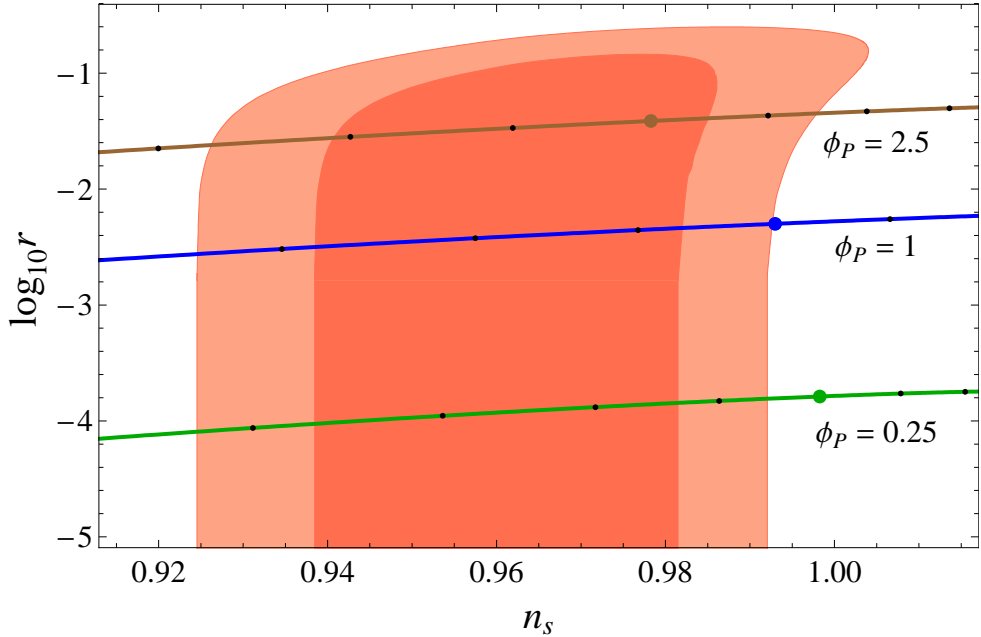} 
\caption{$\log_{10}r$ vs. $n_{s}$ for the cases $\phi_P \equiv \frac{\phi_0}{m_P}= $ 0.25, 1, and 2.5, shown together with the WMAP5+BAO+SN contours.  The smaller black dots denote $f$ values increasing from right to left in increments of 0.05, and the larger colored dots represent inflation starting at the inflection point of the potential, with $f_i\approx 0.60$. With larger values of $\phi_P$, the WMAP5 limit on $r$ can be saturated.} \label{nsr_num}
\end{center}
\end{figure}

Fig. \ref{flogA} shows the behavior of $A$ with respect to both $f$ and $n_s$ in each of our three cases.  Each value of $A$ corresponds to two solutions of $f$ and $n_s$, as expected based on our previous calculations.  Given our current choice of fixed parameters, some values of $A$ result in one hilltop solution and one non-hilltop solution, while others result in two hilltop solutions.  In the numerical case, the value of $f_i$ deviates somewhat from its approximate analytical value $1/\sqrt{3}$, hovering close to 0.60 for all three cases.

The results of our numerical calculations for the RCHI model are compared to WMAP5 in Fig. \ref{nsr_num}.  As $f$ increases along the curves of constant $\phi_P$, $n_s$ decreases to values less than unity.  If $f$ is increased further, these curves pass into the region favored by WMAP5, span the full 2-$\sigma$ range, and pass outside the bounds again.  For the values of $\phi_P$ which we have considered, we obtain a favored range of $f$ corresponding to $0.55\lesssim f \lesssim 0.80$.  Stated differently, the inclusion of suitable radiative correction effects in the HI scenario results in the model becoming more favored by the WMAP5 data.

In the range $0.4\leq f\leq 0.9$ that we have explored, each choice of $\phi_0$ produces a distinct prediction for the size of $r$.  The case $\phi_P=2.5$, which admits trans-Planckian values of the inflaton, results in an appreciable value of $r$ while the others predict a vanishingly small value.  Future missions such as PLANCK may be capable of measuring this quantity with far greater precision.

\section*{Summary}

We obtain a scalar spectral index $n_s$ consistent with WMAP5 data by including fermion-dominated one-loop radiative corrections in a non-susy hybrid inflationary model. A compelling candidate for the origin of these corrections is a Yukawa coupling involving the inflaton and right-handed neutrinos. A red-tilted spectral index is obtained for sub-Planckian values of the inflaton field, which was excluded to 2-$\sigma$ by WMAP5 in the case of the tree level hybrid model. Furthermore, these corrections make accessible both non-hilltop and hilltop-type solutions in this model, and a red-tilted spectrum is achieved primarily for hilltop solutions.

\section*{Acknowledgments}
We thank Nefer {\c S}eno$\breve{\textrm{g}}$uz for valuable discussions. This work is
supported in part by the DOE under grant \# DE-FG02-91ER40626 (Q.S. and M.R.),
by the Bartol Research Institute (M.R.), and by the NSF under grant \# DGE 0538555  (J.W.).


\begin{thebibliography}{99}


%\cite{Komatsu:2008hk}
\bibitem{Komatsu:2008hk}
  E.~Komatsu {\it et al.}  [WMAP Collaboration],
  % ``Five-Year Wilkinson Microwave Anisotropy Probe (WMAP\altaffilmark 1 )
  %Observations:Cosmological Interpretation,''
  arXiv:0803.0547 [astro-ph].
  %%CITATION = ARXIV:0803.0547;%%
  
%\cite{Fukugita:1986hr}
\bibitem{Fukugita:1986hr}
  M.~Fukugita and T.~Yanagida,
  %``Baryogenesis Without Grand Unification,''
  Phys.\ Lett.\  B {\bf 174}, 45 (1986);
  %%CITATION = PHLTA,B174,45;%%
  %\cite{Lazarides:1991wu}
%\bibitem{Lazarides:1991wu}
  G.~Lazarides and Q.~Shafi,
  %``Origin of matter in the inflationary cosmology,''
  Phys.\ Lett.\  B {\bf 258}, 305 (1991), for non-thermal leptogenesis.
  %%CITATION = PHLTA,B258,305;%%


%\cite{NeferSenoguz:2008nn}
\bibitem{NeferSenoguz:2008nn}
  V.~Nefer Senoguz and Q.~Shafi,
  %``Chaotic inflation, radiative corrections and precision cosmology,''
  Phys.\ Lett.\  B {\bf 668}, 6 (2008)
  [arXiv:0806.2798 [hep-ph]].
  %%CITATION = PHLTA,B668,6;%%

%\cite{Linde:1991km}
\bibitem{Linde:1991km}
  A.~D.~Linde,
  %``Axions in inflationary cosmology,''
  Phys.\ Lett.\  B {\bf 259}, 38 (1991);
  %%CITATION = PHLTA,B259,38;%%
%\cite{Linde:1993cn}
%\bibitem{Linde:1993cn}
  A.~D.~Linde,
  %``Hybrid inflation,''
  Phys.\ Rev.\  D {\bf 49}, 748 (1994)
  [arXiv:astro-ph/9307002].
  %%CITATION = PHRVA,D49,748;%%

%\cite{Lazarides:2001zd}
\bibitem{Lazarides:2001zd}
For a review and additional references, see
  G.~Lazarides,
  %``Inflationary cosmology,''
  Lect.\ Notes Phys.\  {\bf 592}, 351 (2002)
  [arXiv:hep-ph/0111328].
  %%CITATION = LNPHA,592,351;%%

%\cite{Clesse:2008pf}
\bibitem{Clesse:2008pf}
  S.~Clesse and J.~Rocher,
  %``Avoiding the blue spectrum and the fine-tuning of initial conditions in
  %hybrid inflation,''
  arXiv:0809.4355 [hep-ph].
  %%CITATION = ARXIV:0809.4355;%%

%\cite{Coleman:1973jx}
\bibitem{Coleman:1973jx}
  S.~R.~Coleman and E.~J.~Weinberg,
  %``Radiative Corrections As The Origin Of Spontaneous Symmetry Breaking,''
  Phys.\ Rev.\  D {\bf 7}, 1888 (1973).  For a review and additional references, see
  %%CITATION = PHRVA,D7,1888;%%
  %\cite{Sher:1988mj}
%\bibitem{Sher:1988mj}
  M.~Sher,
  %``Electroweak Higgs Potentials And Vacuum Stability,''
  Phys.\ Rept.\  {\bf 179}, 273 (1989).
  %%CITATION = PRPLC,179,273;%%


%\cite{Boubekeur:2005zm}
\bibitem{Boubekeur:2005zm}
  L.~Boubekeur and D.~H.~Lyth,
  %``Hilltop inflation,''
  JCAP {\bf 0507}, 010 (2005)
  [arXiv:hep-ph/0502047].
  %%CITATION = JCAPA,0507,010;%%
 
  %\cite{Kohri:2007gq}
\bibitem{Kohri:2007gq}
  K.~Kohri, C.~M.~Lin and D.~H.~Lyth,
  %``More hilltop inflation models,''
  JCAP {\bf 0712}, 004 (2007)
  [arXiv:0707.3826 [hep-ph]];
  %%CITATION = JCAPA,0712,004;%%
  %\cite{Lin:2009yt}
%\bibitem{Lin:2009yt}
 C.~M.~Lin and K.~Cheung,
  %``Reducing the Spectral Index in Supernatural Inflation,''
  arXiv:0901.3280 [hep-ph].
  %%CITATION = ARXIV:0901.3280;%%


%\cite{Liddle:2003as}
\bibitem{Liddle:2003as}
  A.~R.~Liddle and S.~M.~Leach,
  %``How long before the end of inflation were observable perturbations
  %produced?,''
  Phys.\ Rev.\  D {\bf 68}, 103503 (2003)
  [arXiv:astro-ph/0305263].
  %%CITATION = PHRVA,D68,103503;%%

%\cite{Martin:2006rs}
\bibitem{Martin:2006rs}
  J.~Martin and C.~Ringeval,
  %``Inflation after WMAP3: Confronting the slow-roll and exact power  spectra
  %to CMB data,''
  JCAP {\bf 0608}, 009 (2006)
  [arXiv:astro-ph/0605367];
  %%CITATION = JCAPA,0608,009;%%
%\cite{Kolb:1990vq}
%\bibitem{Kolb:1990vq}
% E.~W.~Kolb and M.~S.~Turner,
  %``The Early universe,''
%  Front.\ Phys.\  {\bf 69}, 1 (1990) and references therein.
  %%CITATION = FRPHA,69,1;%%
  A. Linde, {\it Particle Physics and Inflationary Cosmology}
(Harwood Academic Publishers, 1990);
  E.W. Kolb and M.S. Turner, {\it The Early Universe} (Westview,
1990);
  A.R. Liddle and D.H. Lyth, {\it Cosmological
Inflation and Large-Scale Structure} (Cambridge, 2000).
  
  
%\cite{Stewart:1993bc}
\bibitem{Stewart:1993bc}
  E.~D.~Stewart and D.~H.~Lyth,
  % ``A More accurate analytic calculation of the spectrum of cosmological
  %perturbations produced during inflation,''
  Phys.\ Lett.\  B {\bf 302}, 171 (1993)
  [arXiv:gr-qc/9302019].
  %%CITATION = PHLTA,B302,171;%%
  
%\cite{Senoguz:2004vu}
\bibitem{Senoguz:2004vu}
  V.~N.~Senoguz and Q.~Shafi,
  %``Reheat temperature in supersymmetric hybrid inflation models,''
  Phys.\ Rev.\  D {\bf 71}, 043514 (2005)
  [arXiv:hep-ph/0412102] and references therein.
  %%CITATION = PHRVA,D71,043514;%%

%\cite{Dvali:1994ms}
\bibitem{Dvali:1994ms}
 G.~R.~Dvali, Q.~Shafi and R.~K.~Schaefer,
  %``Large scale structure and supersymmetric inflation without fine tuning,''
  Phys.\ Rev.\ Lett.\  {\bf 73}, 1886 (1994)
  [arXiv:hep-ph/9406319];
  %%CITATION = PRLTA,73,1886;%%
%\cite{Copeland:1994vg}
%\bibitem{Copeland:1994vg}
  E.~J.~Copeland, A.~R.~Liddle, D.~H.~Lyth, E.~D.~Stewart and D.~Wands,
  %``False vacuum inflation with Einstein gravity,''
  Phys.\ Rev.\  D {\bf 49}, 6410 (1994)
  [arXiv:astro-ph/9401011];
  %%CITATION = PHRVA,D49,6410;%%
%\cite{Lazarides:1996dv}
%\bibitem{Lazarides:1996dv}
  G.~Lazarides, R.~K.~Schaefer and Q.~Shafi,
  %``Supersymmetric inflation at the grand unification scale,''
  Phys.\ Rev.\  D {\bf 56}, 1324 (1997)
  [arXiv:hep-ph/9608256];
  %%CITATION = PHRVA,D56,1324;%%
%\cite{Linde:1997sj}
%\bibitem{Linde:1997sj}
 A.~D.~Linde and A.~Riotto,
 %``Hybrid inflation in supergravity,''
 Phys.\ Rev.\  D {\bf 56}, 1841 (1997)
 [arXiv:hep-ph/9703209];
 %%CITATION = PHRVA,D56,1841;%%
%\cite{Jeannerot:2000sv}
%\bibitem{Jeannerot:2000sv}
 R.~Jeannerot, S.~Khalil, G.~Lazarides and Q.~Shafi,
 %``Inflation and monopoles in supersymmetric SU(4)c x SU(2)L x SU(2)R,''
 JHEP {\bf 0010}, 012 (2000)
 [arXiv:hep-ph/0002151];
 %%CITATION = JHEPA,0010,012;%%
%\cite{Garbrecht:2006az}
%\bibitem{Garbrecht:2006az}
  B.~Garbrecht, C.~Pallis and A.~Pilaftsis,
  %``Anatomy of F_D-Term Hybrid Inflation,''
  JHEP {\bf 0612}, 038 (2006)
  [arXiv:hep-ph/0605264];
  %%CITATION = JHEPA,0612,038;%%
%\cite{BasteroGil:2006cm}
%\bibitem{BasteroGil:2006cm}
 M.~Bastero-Gil, S.~F.~King and Q.~Shafi,
 %``Supersymmetric hybrid inflation with non-minimal Kaehler potential,''
 Phys.\ Lett.\  B {\bf 651}, 345 (2007)
 [arXiv:hep-ph/0604198];
%\cite{urRehman:2006hu}
%\bibitem{urRehman:2006hu}
  M.~ur Rehman, V.~N.~Senoguz and Q.~Shafi,
  %``Supersymmetric and smooth hybrid inflation In the light of WMAP3,''
  Phys.\ Rev.\  D {\bf 75}, 043522 (2007)
  [arXiv:hep-ph/0612023];
  %%CITATION = PHRVA,D75,043522;%%
%\cite{Pallis:2007du}
%\bibitem{Pallis:2007du}
 C.~Pallis,
  %``Reducing the spectral index in F-term hybrid inflation,''
  arXiv:0710.3074 [hep-ph].
  %%CITATION = ARXIV:0710.3074;%%

%\cite{Rehman:2008qs}
\bibitem{Rehman:2008qs}
  M.~U.~Rehman, Q.~Shafi and J.~R.~Wickman,
  %``GUT Inflation and Proton Decay after WMAP5,''
  Phys.\ Rev.\  D {\bf 78}, 123516 (2008)
  [arXiv:0810.3625 [hep-ph]];
  %%CITATION = PHRVA,D78,123516;%%
%\cite{Kallosh:2007wm}
%\bibitem{Kallosh:2007wm}
 R.~Kallosh and A.~Linde,
  %``Testing String Theory with CMB,''
  JCAP {\bf 0704}, 017 (2007)
  [arXiv:0704.0647 [hep-th]].
  %%CITATION = JCAPA,0704,017;%%



\end{thebibliography}
\end{document}